\documentclass[a4paper,11pt]{article}
\pdfoutput=1 

\usepackage{jinstpub} 

\title{\boldmath Proposal for VEPP-4M beam energy measurement using magnetic spectrometer with Compton calibration and photon detector}

\author{V.V. Kaminskiy,}
\author{N.Yu. Muchnoi and}
\author{V.N. Zhilich}

\affiliation{Budker Institute of Nuclear Physics SB RAS,\\ Novosibirsk, Russia}
\affiliation{Novosibirsk State University, \\Novosibirsk, Russia}

\emailAdd{v.v.kaminskiy@inp.nsk.su}

\abstract{A method for circular e$^-$/e$^+$ accelerator beam energy measurement is proposed. A coordinate of an electron (or a positron) in a focusing magnetic spectrometer built in a circular accelerator depends on its energy, two spectrometer parameters, and the circulating beam energy. The spectrometer parameters can be determined using minimum electron energies from Compton backscattering with two laser wavelengths, and an additional photon coordinate detector. The photon detector is calibrated separately at well-known beam energy. The VEPP-4M collider has appropriate equipment for the method implementation: built-in focusing magnetic spectrometer, Compton calibration system with two lasers and a photon coordinate detector. Thus, the proposed technique could be implemented with minimum efforts; tests and further upgrade are planned.
The beam energy can be defined with expected uncertainty better than $10^{-4}$.
}


\arxivnumber{1705.06527}

\proceeding{Instrumentation for Colliding Beam Physics\\
  27 February - 3 March, 2017\\
  Budker Institute of Nuclear Physics, Novosibirsk, Russia}

\begin{document}
\maketitle
\flushbottom

\section{Motivation}
\label{sec:intro}

Precision bottomonium spectroscopy, $R$ measurement (a total cross-section of $\mathrm{e}^-/\mathrm{e}^+\rightarrow \mathrm{hadrons}$), two-photon processes and other experiments in a beam energy range up to 5.5~GeV are planned at the VEPP-4M e$^-$/e$^+$ collider with the KEDR detector \cite{KEDR} (BINP, Novosibirsk). For this goal beam energy measurement with a relative accuracy better than $10^{-4}$ is needed. The beam energy can be measured in some range with the resonant depolarization technique ($\sim 10^{-6}$ precision) with Touschek polarimeter (operates below 2~GeV) \cite{touschek_polarimeter} and Compton (laser) polarimeter (it is under commissioning to operate above 2~GeV) \cite{laser_polarimeter}. Resonant depolarization technique requires beam polarization, which takes a lot of time. But precise energy measurement is needed in whole energy range with accuracy better than $10^{-4}$ and a minimum measurement time.

In this work a new method of beam energy measurement is proposed. It is based on Compton backscattering and a focusing magnetic spectrometer.

\section{Compton backscattering}

In this paper we imply Compton backscattering as a scattering of a low-energy photon ($\omega_0=1\dots 4$~eV) head on an ultrarelativistic electron (or a positron, $E_0=1.5\dots 5.0$~GeV). If the photon scatters at zero angle (at the initial electron direction), it gains the maximum energy
\begin{equation}
\omega_{\mathrm{max}} = \frac{E_0 \lambda}{1 + \lambda}, \quad \text{where  } \lambda = \frac{4 \omega_0 E_0}{m^2},\,
\end{equation}
where $m$ is electron mass ($c=1$). Hence, the scattered electron has the minimum energy
\begin{equation}
E_{\mathrm{min}} = E_0 - \omega_{\mathrm{max}}
= \frac{E_0}{1 + \lambda}.
\label{eq:emin}
\end{equation}

\section{Focusing magnetic spectrometer}

\begin{figure}[htbp]
\centering
\includegraphics[width=0.6\textwidth]{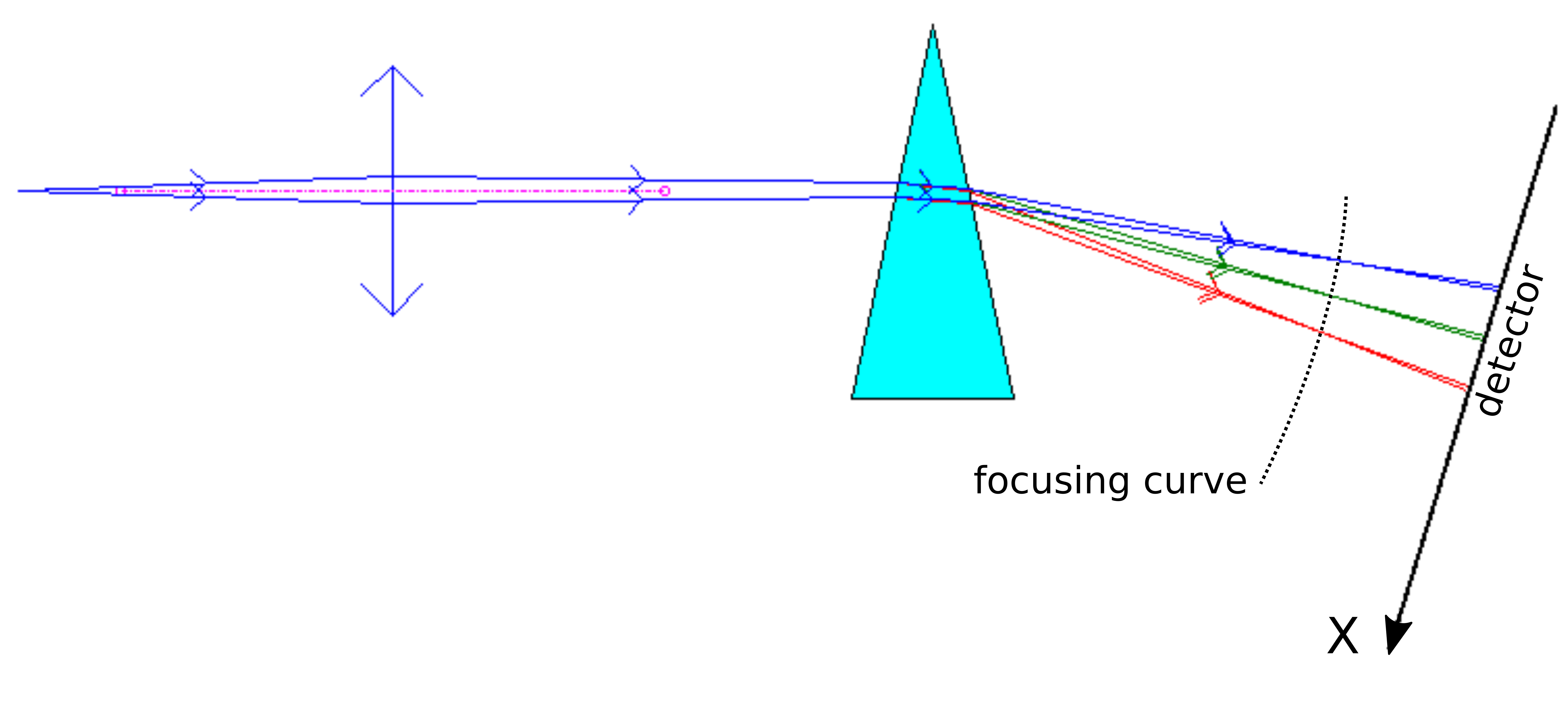}
\caption{Optical analogy of a focusing magnetic spectrometer. Blue rays represent high energy, red rays represent low energies, green rays represent medium energies (color online).} 
\label{fig:analogy}
\end{figure}

A magnetic spectrometer is a device spatially separating fast charged particles by their energies. A dipole magnet is a simplest spectrometer with poorest energy resolution. When provided by a focusing element, the spectrometer becomes a focusing magnetic spectrometer with a drastically improved energy resolution. The principle of the resolution improvement can be illustrated using an optical analogy in Figure~\ref{fig:analogy}. Here an optical prism with anomalous dispersion is equivalent to a bending magnet, an optical lens is equivalent to a quadrupole magnetic lens.

In a linear lattice of an accelerator a coordinate $X$ of an electron after the bending magnet depends on its energy $E$, the circulating beam energy $E_0$ and two parameters $A$ and $B$:
\begin{equation}
X = \frac{A E_0}{E} + B\,.
\label{eq:spectrometer}
\end{equation}

\begin{figure}[htbp]
\centering
\includegraphics[width=\textwidth]{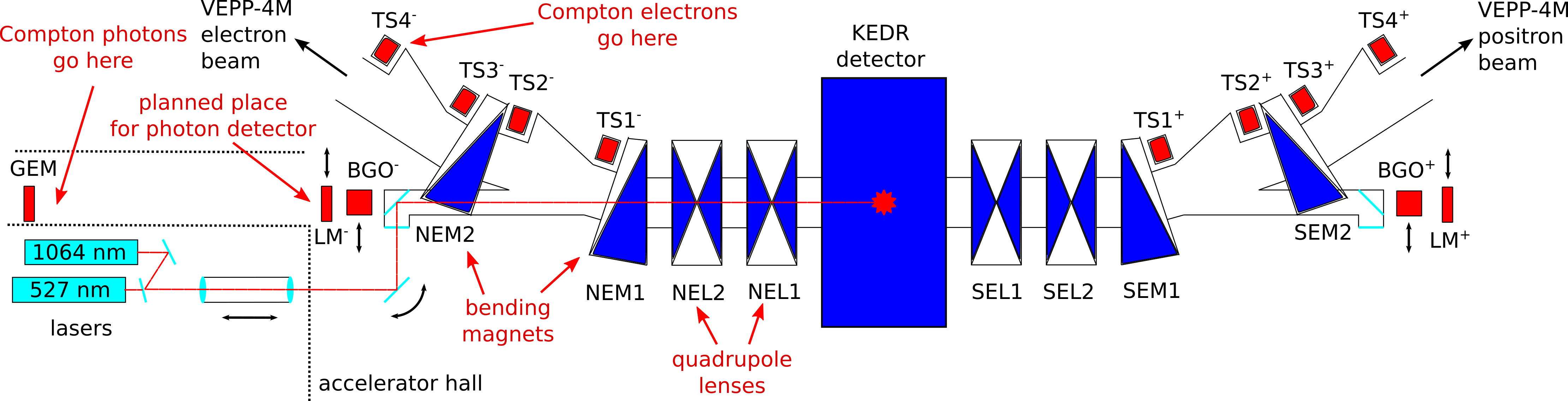}
\caption{Layout of the Tagging System at the KEDR detector.} 
\label{fig:ts}
\end{figure}

The VEPP-4M collider has a focusing magnetic spectrometer embedded into its lattice --- Tagging System \cite{TS} (TS). It is intended for two-photon processes study at the KEDR detector. It has energy resolution for electrons/positrons of $0.05\%\dots0.5\%$ and efficiency for single charged particle (e$^-$ or e$^+$) up to 70\%. The layout of the TS is shown in Figure~\ref{fig:ts}. The TS is located in the experimental section of the VEPP-4M collider (the KEDR detector is in the center); it consist of two symmetrical parts with four quadrupole magnets, (NEL1, NEL2, SEL1, SEL2), four bending magnets (NEM1, NEM2, SEM1, SEM2), four electron coordinate detectors (TS1$^-$, TS2$^-$, TS3$^-$, TS4$^-$) and four positron ones (TS1$^+$, TS2$^+$, TS3$^+$, TS4$^+$).

Compton backscattering is applied for the energy calibration: laser radiation of two wavelengths interacts with the electron/positron beam; scattered electrons/positrons of known energies (see \ref{eq:emin}) are registered by coordinate detectors \cite{TS_calibr}. When the beam energy is precisely measured (currently, only below 2~GeV), two laser wavelengths allow to calibrate the TS4$^{\pm}$ coordinate modules unambiguously and precisely ($10^{-4}$). Figure~\ref{fig:ts} shows the laser system. Laser light of two wavelength are mixed into one beam using a pair of mirrors and focused using motorized beam expander. Then the laser beam is reflected from a movable motorized mirror, inserted in the VEPP-4M vacuum chamber. Being reflected from a mirror, the laser beam interacts with the electron beam in the center of the KEDR detector (e$^-$e$^+$ beam collision point).

Other modules are calibrated with lower precision (1--3\%) using BGO crystal calorimeters and tagging of an electron/positron by a photon emitted in single bremsstrahlung.

\section{Method}

There is a relation between minimum energy of Compton backscattered electron and the energy of the circulating beam~\eqref{eq:emin}. With use of a coordinate detector for Compton photons it allows to define spectrometer parameters and the beam energy simultaneously. 

\subsection{Parameter $B$}
\label{sec:parb}

The parameter $B$ is defined by the circulating beam orbit in the laser-electron interaction point, which is affected by the whole accelerator lattice. Literally, it is the intersection point of the detector plane and the line drawn along the interaction region: $X = B$. Compton photons are emitted along the interaction region. So, if a coordinate of Compton photons were measured in a frame of the electron coordinate detector, the spectrometer parameter $B$ would be determined.

A detector for Compton photons can be placed only downstream the electron detector. To convert the measured coordinate into the electron detector frame, it can be calibrated at once:
\begin{itemize}
\item the beam energy is precisely measured (e.g., by the resonant depolarization);
\item two laser beams with different wavelengths are scattered on the electron beam;
\item with two Compton edges two parameters ($A$ and $B$) are defined (see equations~\eqref{eq:spectrometer} and \eqref{eq:emin}): $X = A (1+\lambda) + B = A + B + 4 A E_0 \omega_0 / m^2$;
\item a constant shift (and possibly, a linear coefficient) for measured Compton photon coordinate is found.
\end{itemize}

\subsection{Parameter $A$}
\label{sec:para}

Parameter $A$ is defined by dispersing and focusing properties of the spectrometer. After scattering two laser beams with wavelengths $\omega_{01}$, $\omega_{02}$ on the electron beam, two Compton edge coordinates are measured. If parameter $B$ were measured, $A$ could be resolved from the following equations:
\begin{gather}
X_{1,2} = A (1+\lambda_{1,2}) + B = A + B + 4 A E_0 \omega_{0(1,2)} / m^2\,, \nonumber \\
E_0 = \frac{X_{1,2} - A - B}{4\omega_{0(1,2)}}m^2 \\
\frac{X_1 - A - B}{\omega_{01}} = \frac{X_2 - A - B}{\omega_{02}} \\
A  = \frac{X_2 \omega_{01} - X_1 \omega_{02}}{\omega_{01} - \omega_{02}} - B.
\end{gather}

\subsection{Energy calculation}

The procedure of the beam energy determination looks as follows:
\begin{itemize}
\item A constant shift for parameter $B$ is defined at once with the precisely measured beam energy, Compton backscattering and photon coordinate detector (see Section~\ref{sec:parb}).
\item Laser radiation of two wavelengths is scattered on the electron beam, $A$ and $B$ are defined by electron and photon coordinate detectors (see Section~\ref{sec:para}).
\item The beam energy is calculated using Compton spectrum edge coordinate $X_{1,2}$ for any laser photon energy $\omega_{0(1,2)}$:
\begin{equation}
E_0 = \left( \frac{X_{1,2} - B}{A} - 1 \right) \frac{m^2}{4 \omega_{0(1,2)}}
\end{equation}
\end{itemize}

\subsection{Uncertainty}
\label{sec:uncertainty}

Relative energy uncertainty of the method proposed is
\begin{equation}
\frac{\delta E_0}{E_0}  = \sqrt{ \left( \frac{\delta X}{X - B} \right)^2 + \left( \frac{\delta B}{X - B} \right)^2 + \left(\frac{\delta A}{A} \right)^2}.
\end{equation}
Here $\delta X$, $\delta A$, $\delta B$ are uncertainties of $X$, $A$, $B$, correspondingly. $\delta X$ is defined by resolution of the electron coordinate detector. $\delta A$ is defined by the spectrometer properties and its calibration (see Sections~\ref{sec:parb} and \ref{sec:para}). $\delta B$ is defined mostly by a coordinate detector for Compton photons and also contains $B$ calibration uncertainty (see Section~\ref{sec:parb}). 

\section{Plans}

With the existing equipment at VEPP-4M the method can be tested in the nearest future.

\begin{figure}[htbp]
\centering
\includegraphics[width=0.7\textwidth]{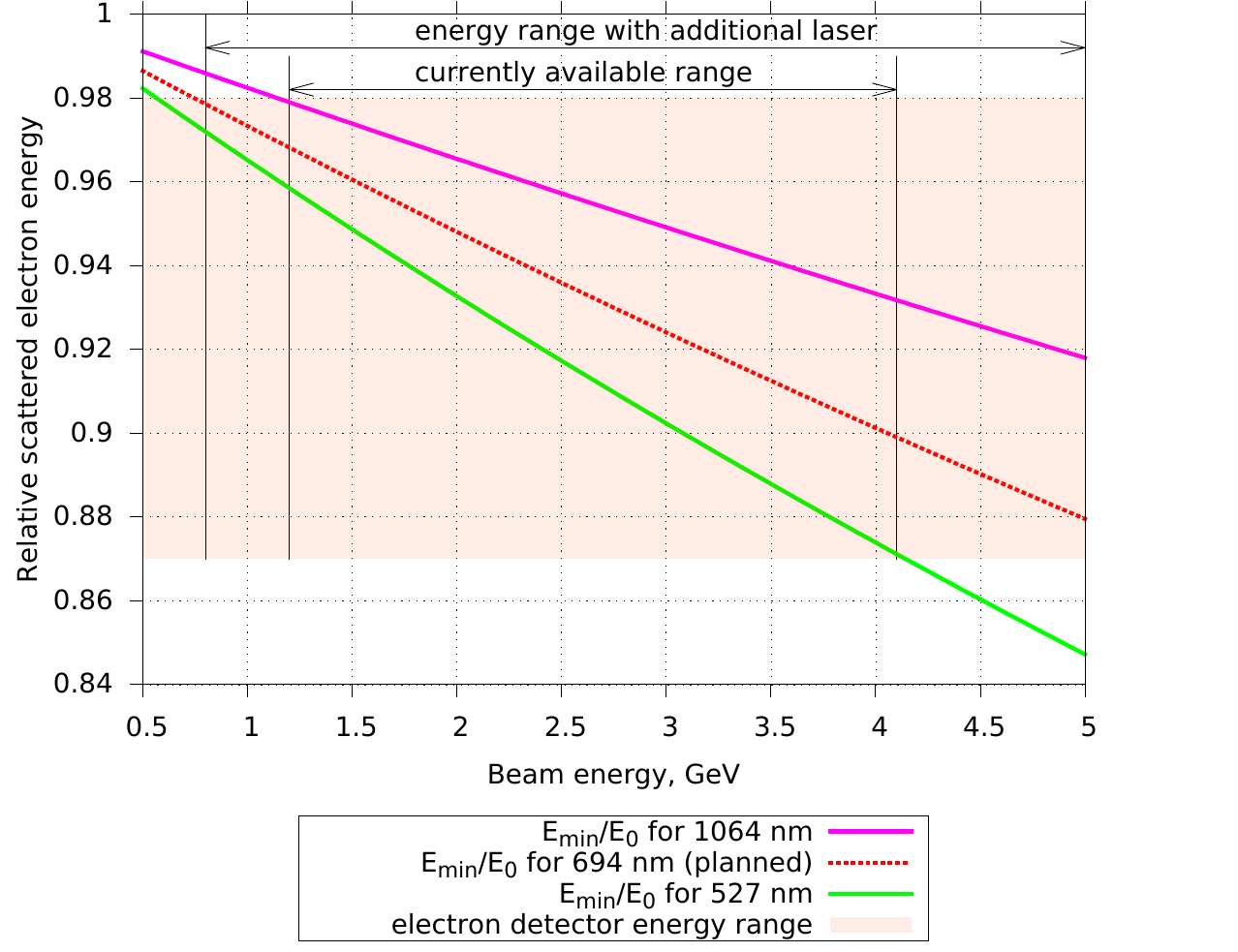} \\
\caption{Possibilities for beam energy measurement (color online). Curves represent minimal energy (relatively to the beam energy) of Compton electrons from different laser wavelengths versus the beam energy (see equation~\ref{eq:emin}). A coloured region represent the energy range of TS4. When Compton edges from two laser wavelengths are in the energy range of TS4, the beam energy can be measured.}
\label{fig:e_w0}
\end{figure}

VEPP-4M has a built-in focusing magnetic spectrometer: the magnetic system and electron coordinate detectors. There is a system for the spectrometer calibration using Compton backscattering: 1064~nm and 527~nm lasers, an optical system, an automated control system. With these lasers the beam energy can be measured in the range of 1.5--4.0~GeV, see Figure~\ref{fig:e_w0}. There is a photon coordinate detector for the "Laser polarimeter" experiment \cite{laser_polarimeter}, located at 18~m downstream from the electron detector.
Coordinate resolution of the TS4$^-$ detector is approximately 100~$\mu$m, which imply a coordinate uncertainty of 50--100~$\mu$m for a coordinate spectrum edge of Compton electrons. The existing photon detector is a gas electron multiplier and has coordinate resolution of 300~$\mu$m, and measured coordinate uncertainty of a Compton photon spot is 40~$\mu$m within reasonable data acquisition time. Spectrometer coefficients $A$ and $B$ for TS4$^-$ are approximately 110~cm. Hence, according to Section~\ref{sec:uncertainty}, estimation of total uncertainty is 10$^{-4}$.

In 2017 tests of the technique are planned at VEPP-4M. Energy measurements by the resonant depolarization will be used for the calibration and uncertainty measurement.

Further improvements can be made:
\begin{itemize}
\item Installing a photon detector (GEM) in front of the luminosity monitor (4~m from the electron detector). It will allow to measure the beam energy in luminosity regime, when the luminosity monitor is used. Being installed at a movable support with stepper motors, it will allow to hold experiments on the VEPP-4M extracted beam. Also it will reduce the uncertainty connected with unknown photon angle (see Sections~\ref{sec:parb} and \ref{sec:uncertainty}).
\item Installing an additional 700~nm laser (e.g., ruby laser) to extend the beam energy measurement range and check the accuracy, see Figure~\ref{fig:e_w0}.
\item Installing an electron detector and data acquisition system with better performance.
\end{itemize}

\section{Conclusion}

A method for the beam energy measurement is proposed. It uses Compton backscattering and detection of scattered electrons in a focusing magnetic spectrometer and scattered photons in a separate photon coordinate detector. The method precision is estimated as $10^{-4}$. Experimental tests and optimization are planned at the VEPP-4M collider.

\bibliographystyle{JHEP}
\bibliography{bem}

\providecommand{\href}[2]{#2}\begingroup\raggedright\begin{thebibliography}{1}

\bibitem{KEDR}
V.~Anashin et~al., \emph{The {KEDR} detector}, {\emph{Physics of particles and
  nuclei} {\bfseries 44} (2013) 657--702}.

\bibitem{touschek_polarimeter}
V.~Blinov et~al., \emph{Beam energy measurements at {VEPP-4M} collider by
  resonant depolarization technique}, {\emph{ICFA Beam Dynamics Newsletter}
  (April, 2009) 181--190}.

\bibitem{laser_polarimeter}
V.~Blinov et~al., \emph{The project of laser polarimeter for beam energy
  measurement of {VEPP-4M} collider by resonance depolarization method},
  {\emph{JINST} {\bfseries 9} (2014) C09010}.

\bibitem{TS}
V.~Aulchenko et~al., \emph{Upgrade of the {KEDR} tagging system}, {\emph{Nucl.
  Instr. \& Meth. A} {\bfseries 494} (2002) 241--245}.

\bibitem{TS_calibr}
V.~Kaminskiy, N.~Muchnoi and V.~Zhilich, \emph{Compton backscattering for the
  calibration of {KEDR} tagging system}, {\emph{JINST} {\bfseries 9} (2014)
  C08021}.

\end{thebibliography}\endgroup

\end{document}